\documentclass{article}

\usepackage{amsfonts}
\usepackage{epsfig}

\begin{document}

\title{\textsc{Comparative Study of Cities as Complex Networks}}

\vspace{1cm}

\author{ D. Volchenkov and Ph. Blanchard
\vspace{0.5cm}\\
{\it  BiBoS, University Bielefeld, Postfach 100131,}\\
{\it D-33501, Bielefeld, Germany} \\
{\it Phone: +49 (0)521 / 106-2972 } \\
{\it Fax: +49 (0)521 / 106-6455 } \\
{\it E-Mail: VOLCHENK@Physik.Uni-Bielefeld.DE}}
\large

\date{\today}
\maketitle

\begin{abstract}

Degree distributions of graph representations for
compact urban patterns are scale-dependent.
Therefore,
the degree statistics alone
does not give us the enough information
 to reach a qualified conclusion on the structure of
  urban spatial
networks.
We investigate the
statistics of far-away neighbors
and propose the new
method for automatic structural
classification of cities.

\end{abstract}

\vspace{0.5cm}

\leftline{\textbf{ PACS codes: 89.75.Fb, 89.75.-k, 89.90.+n} }
 \vspace{0.5cm}

\leftline{\textbf{ Keywords: Complex networks, city space syntax} }

\section{Introduction}
\label{sec:movement}
\noindent

Comparative studies of cities
  adopt classifications, such as
cultural or geographical criteria, and then apply analytical tools
to characterize the existing groups in morphological terms
\cite{Major,Karimi,HillierUDI}. However, in recent space syntax
investigations devoted to the comparative classification
 of urban textures, the
predefined categories have been
avoided and groups have been
interpreted as a result
of the analysis \cite{Meridos,Amorim}.

Methods for automatic
classification or grouping, broadly
termed {\it hierarchical clustering}
have been discussed in \cite{Amorim}.
The general idea
behind the hierarchical clustering
is that elements of any set have
similarities and
differences that can be mapped
as distances in a multi-dimensional
space in which
each characteristic (variable) represents
 an axis.
Then, clusters are created by grouping
isolated elements or subgroups or,
alternatively, splitting the set into smaller
groups, according to the distance between them.

In the present paper, we propose the new 
automatic
classification method based on the
structural statistics of so called justified graphs
used in urban space syntax theory \cite{Hillier96}.

Most real world networks
 can be considered complex by virtue of
 features that do not occur in simple networks.
The encoding of cities into non-planar dual graphs
 (Sec.~\ref{sec:Graphrepresentations})
reveals their complex
structure, \cite{Figueiredo2007}.

If cities were perfect grids in which all lines
 have the same length and number of
junctions, they would be described by
regular graphs exhibiting a
 high level of similarity no matter which part
of urban texture
 is examined. This would create a highly
accessible system that provides multiple routes between any pair of
 locations. It was believed that pure grid systems are easy to
  navigate due to this high
accessibility and to the existence of multiple paths between any
pair of locations \cite{RosvallPRE,RosvallPRL}. Although, the
urban grid minimizes descriptions as long as possible in the ideal
grid all routes are equally probable;
 morphology of the perfect grid does not differentiate main
spaces and movement tend to be dispersed everywhere \cite{Figueiredo2007}.
 Alternatively, if cities were purely hierarchical systems (like trees),
they would clearly have a main space (a hub, a single route between many
pairs of locations) that connects all branches and controls movement
between them.
 This would create a highly segregated, sprawl like system that would
 cause a tough social consequence, \cite{Figueiredo2007}.

However, real cities are neither trees nor perfect grids, but a
combination of these structures that emerge from the social and
constructive processes \cite{Hillier96}. They  maintain enough
differentiation to establish a clear hierarchy \cite{Hanson89}
resulting from a process of negotiation between the public
processes (like trade and exchanges) and the residential process
preserving their traditional structure. The emergent urban network
is usually of a very complex structure which is therefore
naturally subjected to the {\it complex network theory} analysis.

In order to illustrate the applications of complex network theory
methods to the structural investigations of dual graphs representing
 urban environments, we have studied
 five different compact urban patterns.

 Two of them are
situated on islands: Manhattan (with an almost regular grid-like
city plan) and the network of Venice canals
(imprinting the joined
effect of natural, political, and economical factors acting on the
network during many centuries).
In the old city center
of Venice that stretches across 122 small islands in the marshy
 Venetian Lagoon along the Adriatic Sea in northeast Italy,
 the canals serve the function of roads, and every form
 of transport is on water or on foot.

We have also considered two organic cities
founded shortly after the Crusades and developed within the medieval
fortresses: Rothenburg ob der Tauber, the medieval Bavarian city
preserving its original structure from the 13$^\mathrm{th}$ century, and the downtown of
Bielefeld  (Altstadt Bielefeld),
 an economic and cultural center of Eastern Westphalia.

To supplement the study of urban canal networks, we have
investigated that one in the city of Amsterdam. Although
it is not actually isolated from the national canal network, it is
binding to the delta of the Amstel river, forming a dense canal
web exhibiting a high degree of radial symmetry.

\newpage
\begin{center}
{\bf \small Table 1: Some features of studied dual city graphs}

\vspace{0.3cm}

\begin{tabular}{c|c|c|c|c}
   \hline \hline
 Urban pattern   & $N$ & $M$ & $\mathrm{diam}(\mathfrak{G})$ & Intelligibility
    \\ \hline\hline
 Rothenburg ob d.T. & 50 & 115 &  5& 0.85
 \\
Bielefeld (downtown)& 50 & 142 &  6& 0.68
 \\ 
 Amsterdam (canals) & 57 & 200 & 7& 0.91
\\
 Venice (canals) & 96 & 196 &  5& 0.97
 \\ 
  Manhattan & 355 & 3543 &  5& 0.51
 \\
  \hline \hline
\end{tabular}
\end{center}
\vspace{0.3cm}

The scarcity of
physical space is among the most important factors determining the
structure of compact urban patterns. Some characteristics of studied dual city graphs are
given in Tab.1. There, $N$ is the number of open spaces (streets/canals and places)
 in the urban pattern
(the number of nodes in the dual graphs), $M$ is the number of junctions (the number of
edges in the dual graphs);
the graph {\it diameter},
$\mathrm{diam}(\mathfrak{G})$ is the {\it maximal} depth (i.e., the graph-theoretical distance)
between two vertices in a dual graph; the intelligibility parameter
estimates
navigability of the city, suitability for the passage through it \cite{Hillier96}.

In Sec.~\ref{sec:Graphrepresentations}, we discuss the possible
graph representations for the urban spatial networks and explain
our choice of the dual graph representation. In
Sec.~\ref{sec:Urbanmatrix}, we consider the statistics of
justified graphs studied in space syntax theory and introduce the
urban structure matrix describing the statistics of far-away
neighbors. Then, we introduce the structural distance between
cities in Sec.~\ref{sec:cityMetric} and conclude in the last
section.

\section{Graphs and space syntax of urban environments}
\label{sec:Graphrepresentations}
\noindent

Urban space is of rather large scale
 to be seen from a single viewpoint;
maps provide us with its representations by
means of abstract symbols facilitating our
 perceiving and understanding of a city.
 The middle scale and small scale
maps are usually based on Euclidean geometry
 providing spatial objects with precise coordinates
 along their edges and outlines.

There is a long tradition of research articulating urban
environment form using graph-theoretic
principles originating from the paper of Leonard Euler
(see \cite{GraphTheory}).
Graphs have long been regarded as the basic structures
for
representing forms where topological relations are firmly
 embedded within Euclidean
space.
The widespread use of graph theoretic analysis in geographic
 science had been reviewed in \cite{Haggett}
establishing it as central to spatial analysis of urban environments.
In \cite{Kansky}, the basic graph
theory methods had been applied  to the measurements of
 transportation networks.

Network analysis has long been a basic function of geographic
information systems
(GIS) for a variety of applications, in which computational
 modelling of an urban network is based on a graph view in
 which the intersections of linear features are regarded as
nodes, and connections between pairs of nodes are represented
as edges \cite{MillerShaw}.
Similarly, urban forms are usually represented as the patterns
of identifiable urban elements such as
locations or areas (forming nodes in a graph) whose relationships
to one another are often associated with linear
transport routes such as streets within cities \cite{Batty}.
Such planar graph representations define locations or points in
Euclidean plane as nodes or vertices $\{ i\}$, $i=1,\ldots, N$, and
 the edges linking  them together as $i\sim j$, in
 which $\{i,j\}=1,2,\ldots,N.$ The value of a link can
 either be binary, with the value $1$ as $i\sim j$, and
 $0$ otherwise, or be equal to actual physical distance
 between nodes,  $\mathrm{dist}(i,j)$, or to some weight $w_{ij}>0$ quantifying
a certain characteristic property of the link.
We shall call a planar graph representing the Euclidean space
embedding of an urban network as its {\it primary graph}.
 Once a spatial system has been identified and
represented by a graph in this way, it can be subjected to
the graph theoretic analysis.

A {\it spatial network} of a city is a network of the spatial
 elements of urban environments. They
are derived from maps of {\it open spaces} (streets, places, and roundabouts).
Open spaces may be broken down into components; most simply, these
might be street segments, which can be linked into a network via
their intersections and analyzed as a networks of {\it movement
choices}. The study of spatial configuration is instrumental in
predicting {\it human behavior}, for instance, pedestrian
movements in urban environments \cite{Hillier96}. A set of
theories and techniques for the analysis of spatial configurations
is called {\it space syntax} \cite{Jiang98}. Space syntax is
established on a quite sophisticated speculation that the
evolution of built form can be explained in analogy to the way
biological forms unravel \cite{SSyntax}. It has been developed as
a method for analyzing space in an urban environment capturing its
quality
 as being comprehendible and easily navigable \cite{Hillier96}.
Although,  in its initial form, space syntax was focused mainly on
patterns of pedestrian movement in cities, later the  various
space syntax measures of urban configuration had been found to be
correlated with the different aspects of social life,
\cite{Ratti2004}.

Decomposition of a space map into a complete set of
intersecting axial lines,  the fewest and
longest lines of sight that pass through every open space comprising any system,
produces an axial map or an overlapping convex map respectively.
Axial lines and convex spaces may be treated as the {\it spatial elements}
 (nodes of a morphological graph),
 while either the {\it junctions} of axial lines or the {\it overlaps} of
 convex  spaces may be considered as the edges linking  spatial elements
 into a single  graph unveiling the
topological relationships  between all open elements of the urban space.
In what follows,  we shall call this morphological representation of urban network
as a {\it dual graph}.

The transition to a dual graph is a topologically non-trivial
transformation of a planar primary graph into a non-planar one which
encapsulates the hierarchy and structure of the urban area and also corresponds
 to perception of space that people experience when
travelling along routes through the environment.

In Fig.~1, we have presented the glossary establishing a correspondence
 between several typical elements of urban environments and the certain subgraphs
 of dual graphs.
The dual transformation replaces the 1D open segments (streets) by the
 zero-dimensional nodes, Fig.~1(1).
\begin{figure}[ht]
\label{F1_11}
 \noindent
\begin{center}
\begin{tabular}{llrr}
 1. &\epsfig{file=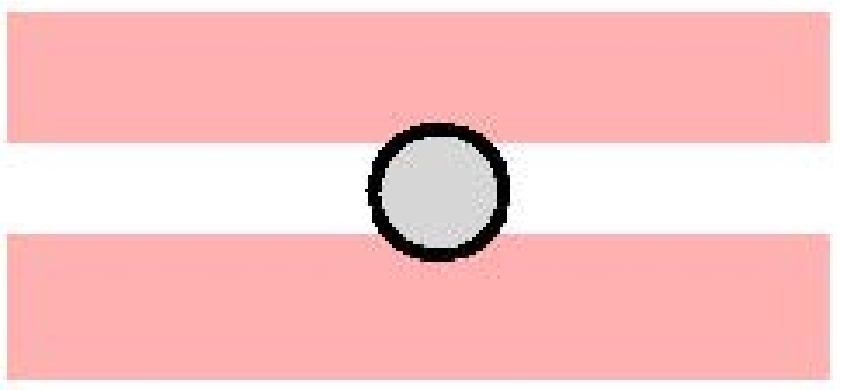, width=3.0cm, height =1.5cm}&2.&
 \epsfig{file=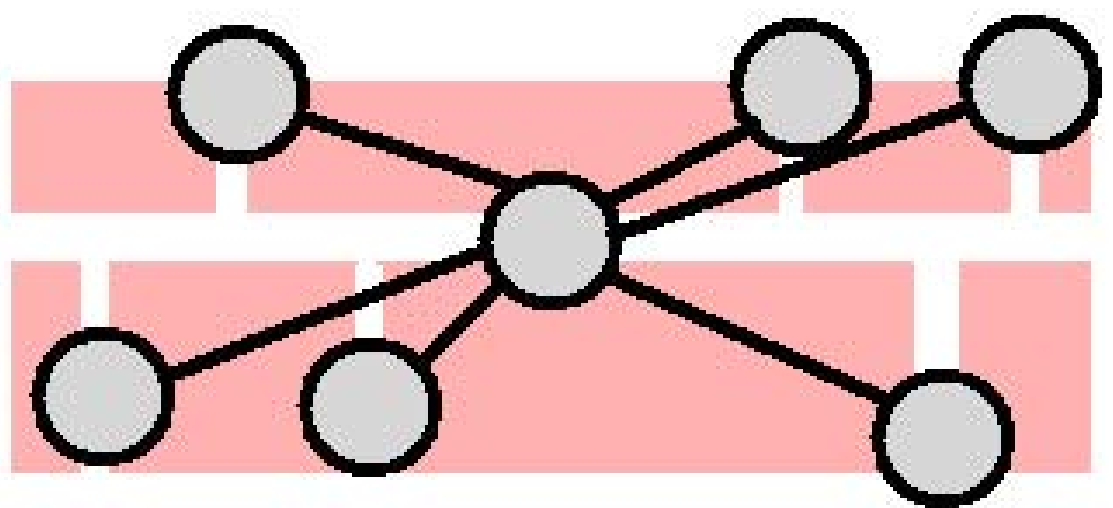, width=3.0cm, height =1.5cm} \\
 3. &\epsfig{file=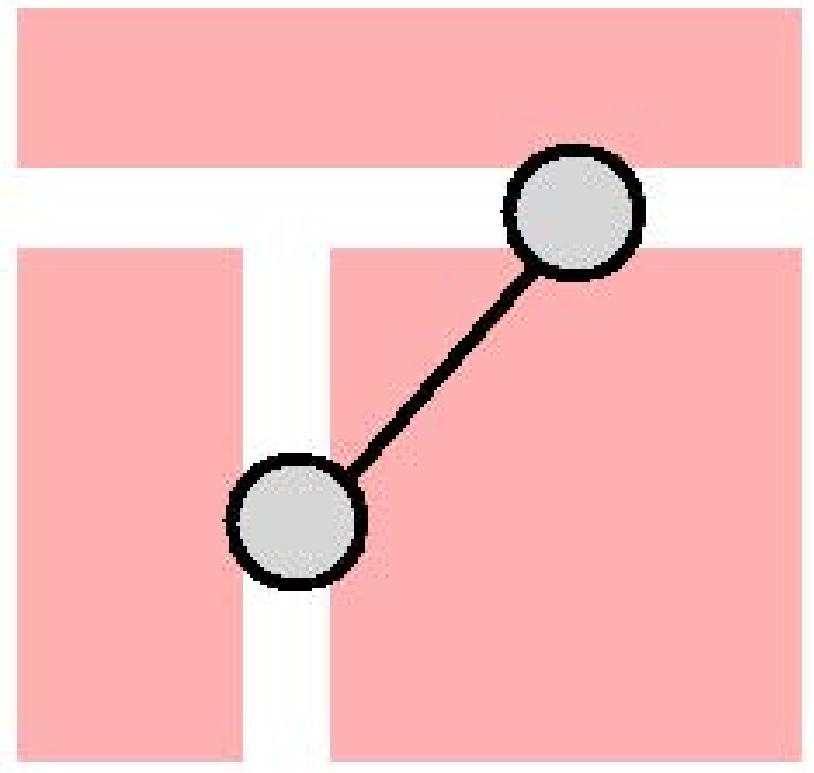,width=3.0cm, height =3.0cm}&4.&
 \epsfig{file=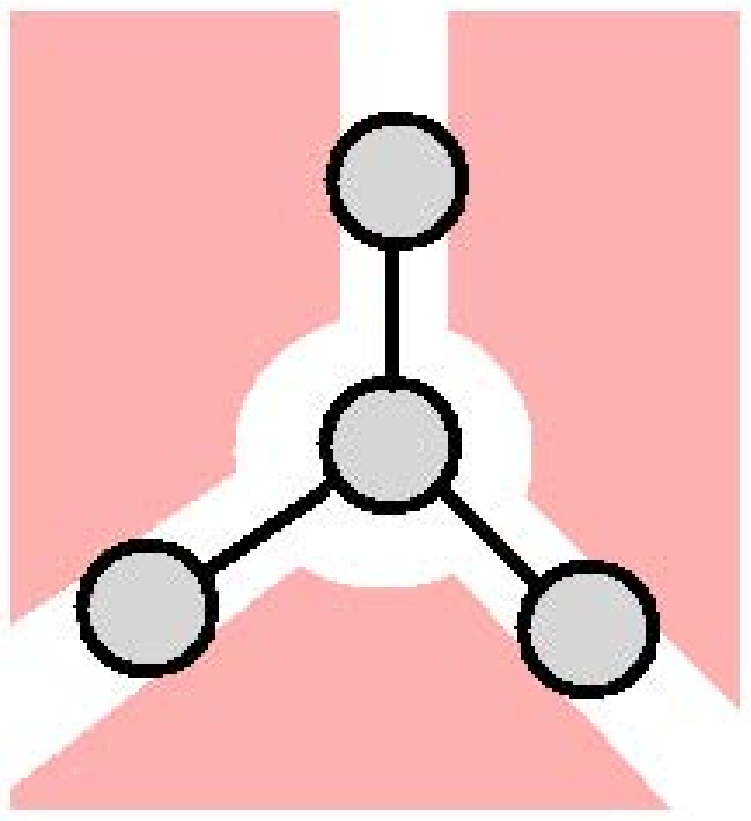, width=3.0cm, height =3.0cm} \\
 5. &\epsfig{file=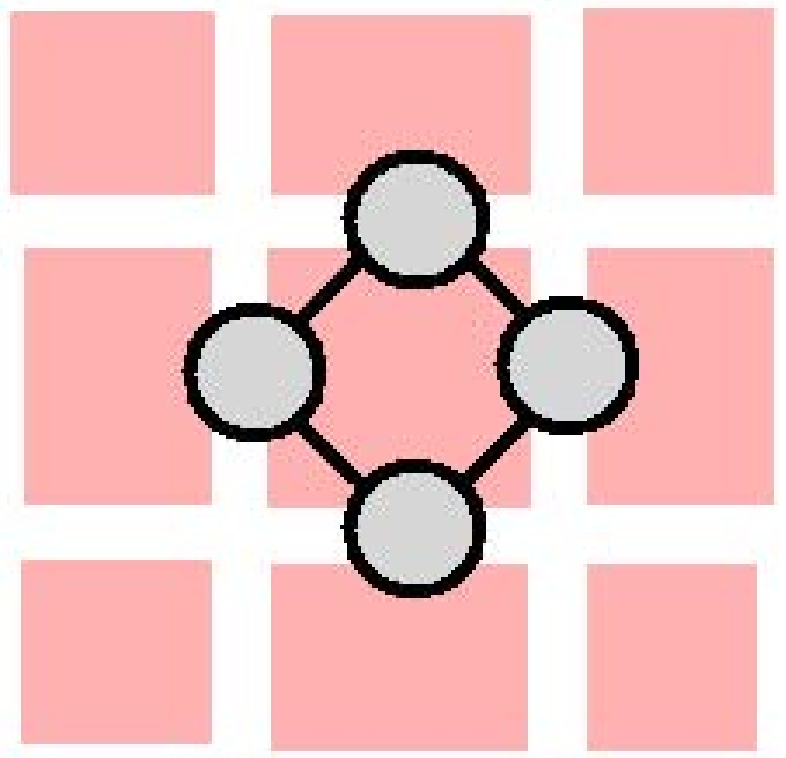,width=3.0cm, height =3.0cm}&6.&
 \epsfig{file=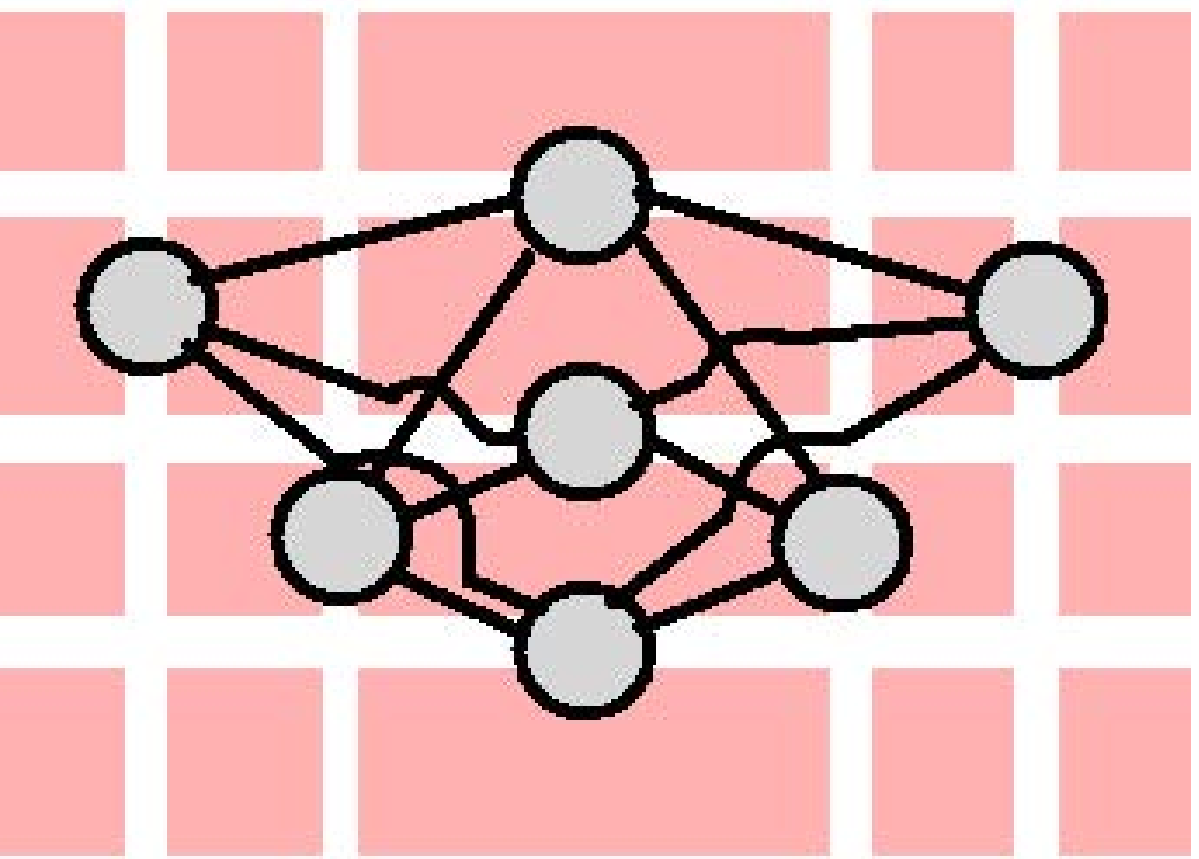, width=3.0cm, height =3.0cm} \\
\end{tabular}
\end{center}
\caption{The dual transformation glossary between the typical elements of
urban environments and the certain subgraphs of dual graphs.}
\end{figure}

The sprawl like developments consisting of a number of blind passes branching
 off a main route are changed to the star subgraphs having a hub and a number
 of client nodes, Fig.~1(2). Junctions and crossroads are replaced with
 edges connecting the corresponding nodes of the dual graph, Fig.1(3).
  Places and roundabouts are considered as the independent topological
  objects and acquire the individual IDs being nodes in the dual
  graph Fig.~1(4). Cycles are converted into  cycles of
  the same lengthes Fig.~1(5). A regular grid pattern
  is shown in Fig.~1(6). Its dual graph representation
  is called a {complete bipartite} graph, where the set of vertices
  can be divided into two disjoint subsets  such that no edge has
   both end-points in the same subset, and every line joining the
   two subsets is present, \cite{Krueger89}.
These sets can be naturally interpreted as those of the vertical
and horizontal edges in the primary graphs (streets and avenues).
In bipartite graphs, all closed paths are of even length, \cite{Skiena}.

 It is the dual
graph transformation which allows to separate the effects of order
and of structure while analyzing a transport network on the
morphological ground. It converts the repeating geometrical
elements expressing the order in the urban developments into the
{\it twins nodes}, the pairs of nodes such that any other is
adjacent either to them both or to neither of them. Examples
of twins nodes can be found in Fig.~1(2,4,5,6).

\section{Justified graphs and urban structure matrices: statistics of far away neighbors}
\label{sec:Urbanmatrix}
\noindent

In space syntax theory,  graph-based models
 of space are used in order
to investigate the influence of the
shape
and configuration of environments on
human spatial behavior and experience.
A number of configurational {\it measures}
have been introduced in so far
in quantitative representations
of relationships between
spaces of urban areas and buildings.
Although similar parameters
quantifying
connectivity and centrality  of nodes in a graph
have been independently invented and
extensively studied during
the last century in a varied range of disciplines
including
computer science, biology, economics, and sociology, the
syntactic measures are by no means just the new names for the well
known quantities.
In space syntax, the spaces are
understood as voids between
buildings restraining
traffic.

The main focus of the space syntax study is on the
relative proximity (or {\it accessibility}) between different locations
which involves calculating graph-theoretical {\it distances} between
nodes of the dual graphs and associating these distances with densities and
intensities of human activity which occur at different open spaces and
along the links which connect them \cite{Hansen59,Wilson70,Batty}.

Space {\it adjacency} is a basic rule to form axial maps:
two axial lines
 intersected are regarded as adjacency.
The dual
graph $\mathfrak{G}$ is  a labelled graph, in which labels are
assigned to its nodes representing open spaces of the system.
Two spaces, $i$ and $j$, are
held to be {\it adjacent} in the dual graph $\mathfrak{G}$
when it is possible
to {\it move freely} from one
space to another, without passing through any
intervening. These two nodes, each node representing
 an individual vista space, are connected by an
edge of the dual graph.
A {\it syntactic step} is defined as the direct connection $i\sim j$
 between neighboring individual open spaces
or between overlapping isovists, $i,j\in \mathfrak{G}$.

Although graphs are usually shown diagrammatically, this
is only
reasonable when the number of vertices and edges
is small enough.
Graphs can also be represented in the form of matrices.  The major
advantage of matrix representation is that the calculation of paths
 and cycles can easily be performed using well known operations on
  matrices.
The {\it adjacency matrix}  ${\bf A}_\mathfrak{G}$ of the dual graph
   $\mathfrak{G}$ is defined as follows:
\begin{equation}
\label{adjacencymatrix}
({\bf A}_\mathfrak{G})_{ij}=\left\{
\begin{array}{ll}
1,& i\sim j, \\
0,& \mathrm{otherwise}.
\end{array}
\right.
\end{equation}
Let us note that
rows and columns of ${\bf A}_\mathfrak{G}$
corresponding to
 the twins nodes are identical.
{\it Depth} is a topological distance
between nodes
in the dual graph $\mathfrak{G}$.
Two open spaces,  $i$  and $j$, are said to
 be at  depth $d_{ij}$ if
the
{\it least number} of syntactic steps
needed to reach one node from the other
is $d_{ij}$, \cite{glossary}.
In particular,
if two axial lines intersect
on the axial map, then the depth between them
 is equal to one.
Given a connected dual graph, we can compute
the depths between all pairs of its nodes.

In order to visualize
 the {\it overall depth}
of spatial layout seen from one of its points, the {\it
justified graph} is used in
 the space syntax analysis, \cite{Hillerhanson}.
Justified graph is a graph restructured so
that a specific space is placed at the bottom,
the {\it root space}. All spaces one syntactic step
away from root space are put on the {\it first level}
above, all spaces two steps away on the
{\it second level}, etc, \cite{glossary}.
  A tree-like justified
graph has most of the nodes many steps (levels)
away from the bottom node. In such a system
the mean depth is high and described as {\it deep}.
A bush-like justified graph has most of the
nodes near the bottom and the system is
described as {\it shallow}.

We begin with the analysis of
 a dual graph representing a piece of
urban texture with a selection of a street (a node in the dual
graph) as a starting point. This street will be intersected by a
number, $n_{d=1}$, of other streets, which are labelled depth
$d=1$. These $n_{d=1}$ streets will then be intersected by
$n_{d=2}$ streets, which are labelled depth $d=2$, etc. As a
consequence of the analysis, each street in the map is numbered
according to how many changes of direction (or elementary
navigation actions) separate it from the starting street.

\begin{figure}[ht]
 \noindent
\begin{center}
\epsfig{file=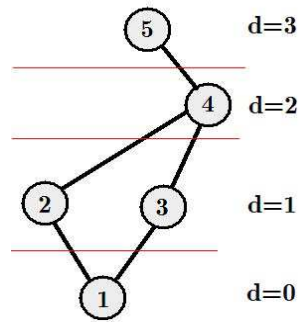, angle=0,width =4cm, height =4.5cm}
  \end{center}
\caption{\small An example of the justified graph with the root node $\# 1$.}
\label{Fig1_18}
\end{figure}

In space syntax theory, the centrality level of a node is analyzed by means
of a {\it justified
graph} restructured so
that the node is placed
at the bottom being a root of the graph, and
all spaces (nodes) one syntactic step
away from the root space are put on
the first level
above, all spaces two spaces away on the
second level, etc.
 In Fig.~\ref{Fig1_18}, we have presented a small
graph of five nodes, with the node $\# 1$ being
 a root.

If a root node belongs to a tree-like component of the dual graph,
it obviously has most of the nodes many steps away. Such a "deep"
space  with a  high mean depth is segregated from most of other
spaces in the city probably  belonging to a city fringe. Being
chosen as roots, the nodes of high centrality produce shallow
justified graphs with most of nodes near the bottom.

\begin{figure}[ht]
 \noindent
\begin{center}
\epsfig{file=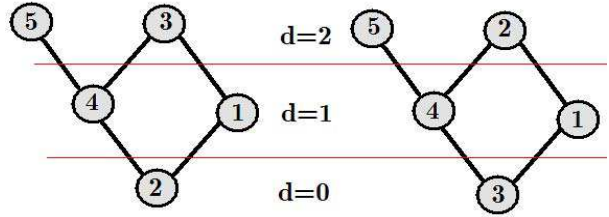, angle=0,width =8cm, height =3cm}
  \end{center}
\caption{\small The justified graphs drawn with the twins
 nodes taken as the roots are identical.}
\label{Fig1_19}
\end{figure}

Given a root node $i$, the {\it justified graph} can be specified by the vector
$\left( n_1, n_2,\ldots n_{\mathrm{diam}(\mathfrak{G})}\right)_i$, in which  $n_d$ is
the number of nodes being at a depth $d$ away from the root node $i$.
For example, the specification vector for the justified graph given in
 Fig.~\ref{Fig1_18} is $(2,1,1)_1$ where we put the index $(\ldots)_1$
 indicating that the graph has been drawn with respect to the
 root node $\# 1$.
Other four justified graphs which can be constructed from that
 one shown in Fig.~\ref{Fig1_18} by placing other nodes at the bottom
are specified by the vectors $(2,2,0)_2,$ $(2,2,0)_3,$
$(3,1,0)_4,$ $ (1,2,1)_5.$
Let us note that the specification vectors correspondent
to the twins nodes $\# 2$ and $\# 3$ are identical as well
as the relevant justified graphs (see Fig.~\ref{Fig1_19}).

It is worth to mention that the first component $n_1$ of the
specification vector $\left( n_1, n_2,\ldots n_{\mathrm{diam}
(\mathfrak{G})}\right)_i$ for the node $i\in\mathfrak{G}$ is
 nothing but its connectivity, $\mathrm{Connectivity}(i)=n_1$.
  It is also clear that specification vectors generalize the notion of connectivity
accounting for all far away neighbors a given node has at a
distance $1\leq d \leq \mathrm{diam}(\mathfrak{G})$ away from it.

We introduce the automatic classification method for
urban textures based on
the statistics over all specification vectors
$\left( n_1, n_2,\right.$ $\left.\ldots n_{\mathrm{diam}(\mathfrak{G})}\right)$
for a given dual graph.

Given a dual graph $\mathfrak{G}$, we introduce a rectangular,
 $\mathrm{diam}(\mathfrak{G})\times N$ integer
 {\it structure matrix} ${\bf J}$, which entries, $J_{nd}$,
 equal to the numbers of specification vectors, with $d$-component
  $n_d = n$. In terms of graph theory, $J_{nd}$ is nothing else
  but the {\it number of nodes} in $\mathfrak{G}$ which have
  precisely $n$ neighbors at a distance $d$ away from them.
By definition the matrix ${\bf J}$ is independent of
a particular layout of street IDs being inherent to the given $\mathfrak{G}$.
The similar matrix has been recently proposed in \cite{Portraits}
in order to portrait  large complex networks.

\begin{figure}[ht]
 \noindent
\begin{center}
\epsfig{file=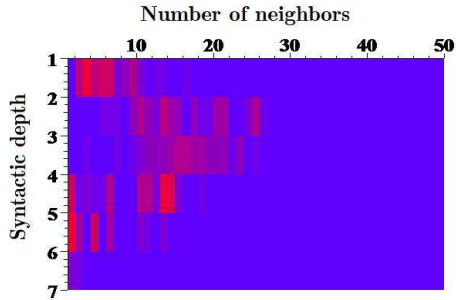, angle=0,width =6cm, height =4cm}
  \end{center}
\caption{\small The urban structure
matrix of the Bielefeld downtown shown in the   linear color scale.
 The matrix elements, $J_{nd}$, are the numbers
of spaces in the city which have precisely
 $n$ neighbors $d$ syntactic steps away
(large values are brighter: red color corresponds to 10 spaces; blue is for zero).}
\label{Fig1_20}
\end{figure}
Despite
its strong influence on various properties of complex networks, the degree
statistics is but one of many important characteristics of a graph.
Complex networks may possess similar degree distributions
yet differ widely in other properties.
The urban structure matrix  ${\bf J}$ contains more detailed information on the
structure of the graph $\mathfrak{G}$ than the probability degree distribution alone.
The degree distribution statistics  (the statistics
 of immediate neighbors)
is encoded in the first row of the urban structure matrix ${\bf J}$ (at $d=1$),
\begin{equation}
\label{degdistrJ}
J_{1,k}=NP(k), \quad P(k)=\Pr\left[i\in\mathfrak{G}|
\deg(i)=k\right].
\end{equation}
Other rows of ${\bf J}$ describe the statistics of far away
neighbors a node has at a distance $d$,
\begin{equation}
\label{rowdistrJ}
J_{dk}=N\cdot P_d(k),\quad P_d(k)=\Pr\left[i\in\mathfrak{G}|(n_d)_i=k\right].
\end{equation}
In Fig.~\ref{Fig1_20}, the matrix plot of $J_{nd}$ for the
downtown of Bielefeld is sketched out. It is interesting to note
that the distributions
of far-away neighbors ($d>1$)
are far from being the monotonous decreasing
functions of the number of neighbors,
 in contrast to the distribution of immediate neighbors,
$P(k)=\Pr\left[i\in\mathfrak{G}|
\deg(i)=k\right]$.

The matrix plot (\ref{Fig1_20}) unveils
a complicated, scale
dependent structure of the
urban texture.
The urban structure
matrix can be used in order to define a
structural distance
between the different urban textures.

\section{Cumulative structure matrix and structural distance between cities}
\label{sec:cityMetric}
\noindent

The statistics of far away neighbors can be presented in the form
of cumulative distribution functions $\mathfrak{P}_d(n)$
quantifying the probabilities that the number of neighbors a node
has at a distance $d$ is greater than or equal to $n$. Being
monotonous functions of $n$,  cumulative distributions reduce the
noise in the distribution tails, however the adjacent points on
their plots are not statistically independent.

The cumulative distributions $\mathfrak{P}_d(n)$
are encoded in rows of the {\it cumulative
 structure matrix},
\begin{equation}
\label{CumulativeMatrix}
\mathfrak{J}_{dn}=\frac{\sum_{k\leq n}^N{J_{dk}}}{\sum_{k=1}^N{J_{dk}}},
\quad k\geq n.
\end{equation}
The plots of cumulative distributions of far-away neighbors in the
dual city graphs shown
in Fig.~\ref{Fig1_21}-\ref{Fig1_21a}.

\subsection{Portraits of cities as complex networks}

\begin{figure}[ht]
 \noindent
\begin{center}
\epsfig{file=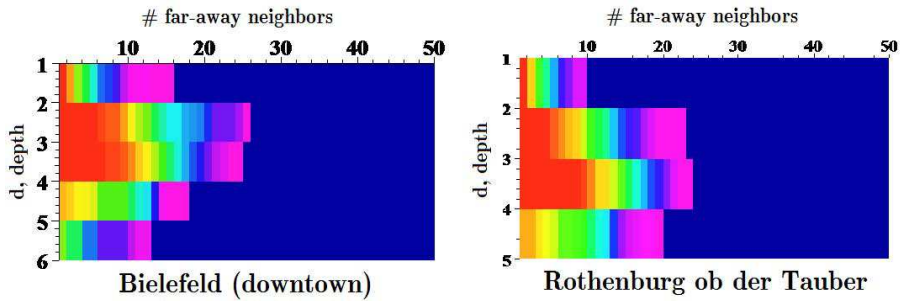, angle=0,width =12cm, height =4cm}
  \end{center}
\caption{\small The cumulative distributions of far-away neighbors
in the dual graphs of German medieval cities. The cumulative
degree distribution are shown in the first row. The cumulative
distributions of far-away neighbors are encoded in the second and
forthcoming rows. Probability is ranked from zero (dark blue) to 1
(red).} \label{Fig1_21}
\end{figure}

Usual networks constitute graphs in which most nodes are neighbors
of one
 another, and
the number of far-away neighbors grows up fast with the distance apart
 from a node. The portraits of German medieval
cities are the examples (Fig.~\ref{Fig1_21}).

\begin{figure}[ht]
 \noindent
\begin{center}
\epsfig{file=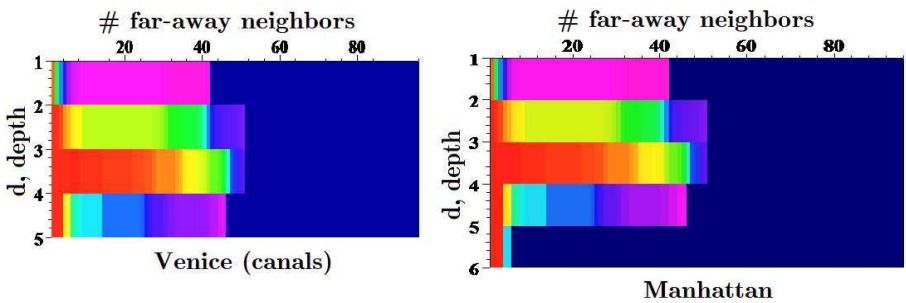, angle=0,width =12cm, height =4cm}
  \end{center}
\caption{\small The cumulative distributions of far-away neighbors
in the dual graphs of Venetian canal network and of the street grid in Manhattan.
Probability is ranked from zero (dark blue) to 1
(red).}
\label{Fig1_21a}
\end{figure}

It is seen from the red profile on Fig.~\ref{Fig1_21} that most of streets
 in these organic cities are just by three syntactic steps away from each
  other. In contrast to them, the cumulative distributions of far-away
  neighbors
in the dual city graphs displayed in Fig.~\ref{Fig1_21a} show that while
 for the  street grid in Manhattan and for the canal network of Venice
most nodes can still be reached in at most three syntactic steps, there
 are also a noticeable fraction of relatively segregated
  nodes few steps deeper than the others.

The structure of cumulative distributions shown in
Fig.~\ref{Fig1_21a} evidences in favor of that the correspondent
dual graphs have high representation of {\it cliques}, and
subgraphs that miss just a few edges of being cliques. In complex
network theory, this phenomenon is referred to as a {\it small
world}. Small-world networks are characterized by a high clustering
coefficient having connections between almost any two nodes within
them. Hubs - nodes in the network with a high number of
connections serving as the common connections mediating the short
path lengths between other edge are commonly associated with
small-world networks, \cite{Buchanan}.

The small world phenomenon indicated on the city portraits in
Fig.~\ref{Fig1_21a} acquires an elegant
 explanation in the framework of space syntax theory.

In \cite{SSBeijing}, \cite{Hillier96}, it has been suggested that
the spatial structure of organic cities is shaped by the public
processes ordered in such a way as to maximize the presence of
people in the central areas. In such a context, the {\it compact}
structure of German medieval burghs (Fig.~\ref{Fig1_21}) uncovers its
historical functional pertinence.

The tendency to shorten syntactic distances in  urban
space networks induced by the public processes
 is complemented in the "small world" cities
(like those portrayed in Fig.~\ref{Fig1_21a})
with the residential process  which shapes relations between
inhabitants and strangers preserving the original residential
culture against unsanctioned invasion of privacy. While most of
streets and canals characterized with an excellent accessibility
promotes commercial activities and intensifies cultural exchanges,
the certain districts of such cities stipulate the alternative
tendency having the residential areas relatively segregated from
the rest of the urban fabric.

\subsection{Structural distance between cities}

A simple metric comparing two urban patterns represented by their
 dual graphs, $\mathfrak{G}$ and $\mathfrak{G}'$,  may be
 defined, using their cumulative structure matrix,
  $\mathfrak{J}$ and $\mathfrak{J}'$.

In order to measure dissimilarities between distributions,
the {\it Wasserstein distance} is widely used
in probabilistic contexts \cite{Rachev}.
When applied to sample distributions the Wasserstein
distance leads to the $L_2$-based tests of goodness of fit.
Given two sets of probabilities, $P_1$ and $P_2$, with finite
 second order moment, the Wasserstein distance between them
is defined as the lowest $L_2$-distance between random
variables $X_1$ and $X_2$ defined in any probability space, with these
distribution laws:
\begin{equation}
\label{Wasserstein}
W(P_1,P_2)=\inf\left\{\sqrt{E(X_1-X_2)^2},{\ } X_1\sim P_1, X_2\sim P_2\right\},
\end{equation}
in which $E$ is the mean.
It is worth to mention that the Wasserstein distance
 is not the only possible measure of dissimilarities
  between distributions which can be implemented in the
  comparative analysis of urban structures. For example,
  in \cite{Portraits} the Kolmogorov-Smirnov test
   \cite{Conover} has been used instead of the
   Wasserstein distance (\ref{Wasserstein}).

Motivated by the Wasserstein distance \cite{Rachev}, we
 introduce the following statistic between the
 corresponding pairs of rows $\mathfrak{J}_d$ and
 $\mathfrak{J}'_d$,
$1\leq d\leq \min (\mathrm{diam}(\mathfrak{G}),$ $\mathrm{diam}(\mathfrak{G}'))$:
\begin{equation}
\label{W_d}
W_d= \sqrt{ \frac{\sum_{n=1}^{\min(N,N')}(\mathfrak{J}_{dn}-
\mathfrak{J}'_{dn})^2}{\min(N,N')}},
\end{equation}
where $N$ and $N'$ are the sizes of the dual graphs
 $\mathfrak{G}$ and $\mathfrak{G}'$ respectively.

It is clear that while comparing the dual graphs of different
sizes,  we use the "upper left corners" of their cumulative
structure matrices suggesting that near neighbors located just a
few steps away have greater impact on network properties. For
instance, only the  shallow neighbors  have been taken into
account in the traditional space syntax analysis of urban
textures. Therefore, it seems reasonable to assign weights to the
shallow neighbors statistics  while estimating their contributions
into similarity /dissimilarity of cities. An appropriate set of
weights for distances has been proposed in \cite{Portraits},
\begin{equation}
\label{setweights}
\alpha_d=\sum_{n=1}^{\min(N,N')}(J_{dn}+J'_{dn}).
\end{equation}
Then, the {\it structural distance} between two dual graphs,
$\mathfrak{G}$ and $\mathfrak{G}'$, is calculated by:
\begin{equation}
\label{scalar_distance}
\Delta(\mathfrak{G},\mathfrak{G}')=
\frac{\sum_{d=1}^{\min \left(\mathrm{diam}
(\mathfrak{G}),\mathrm{diam}(\mathfrak{G}')\right)}\alpha_d W_d}
{\sum_{d=1}^{\min \left(\mathrm{diam}(\mathfrak{G}),\mathrm{diam}
(\mathfrak{G}')\right)}\alpha_d}.
\end{equation}
It equals zero if the distributions of $\min(N,N')$ far-away
neighbors over the first $\min
\left(\mathrm{diam}(\mathfrak{G}),\mathrm{diam}(\mathfrak{G}')\right)$
shells of two dual graphs, $\mathfrak{G}$ and $\mathfrak{G}'$, are
identical. Alternatively, the structural distance approaches one
as these distributions are orthogonal.

\begin{center}
{\bf \small Table 2: The table of structural distances
 between the dual graphs of compact urban patterns.}

\vspace{0.3cm}

\begin{tabular}{c||c|c|c|c|c}
   \hline \hline
 & Bielefeld   & Rothenburg &  Amsterdam & Venice & Manhattan
    \\ \hline\hline
 Bielefeld & 0 & 0.043 &  0.321& 0.139& 0.396
 \\ \hline
Rothenburg& 0.043 & 0 &  0.127& 0.324 & 0.412
 \\ \hline
 Amsterdam& 0.139 & 0.127 & 0& 0.316 & 0.450
\\ \hline
 Venice  & 0.321 & 0.324 &  0.316 & 0 & 0.244
 \\ \hline
  Manhattan & 0.396 & 0.412 &  0.450& 0.244&0
 \\ \hline
  \hline
\end{tabular}
\end{center}

\vspace{0.3cm}

In Tab.~2, we have presented the structural
 distances between the dual graphs of five compact urban
 patterns calculated in accordance to (\ref{scalar_distance}).

It is important to note that the structural distances given in the
Tab.~2 have been  calculated independently for each pair of urban
textures with reference to their sizes and distributions of far
away neighbors. It is obvious that they do not belong to the same
space and therefore cannot be immediately compared.

\section{Discussion and Conclusion}
\label{Discussion}
\noindent

The
{\it degree distribution} has become an important
 concept in complex network theory
 describing the topology of complex networks.
It originates from
 the study of random graphs by Erd\"{o}s and R\'{e}nyi, \cite{ErdosRenyi}.
The importance of the implemented
street identification principle is worth
 mentioning for the investigations on the degree
statistics of dual city graphs.
The comparative investigations
of different street patterns performed in \cite{Cardillo,Porta}
 implementing
the ICN principle reveal scale-free degree
distributions for the vertices of dual graphs.
However, in \cite{Jiang2004}
it has been reported that under the street-name approach the
dual graphs exhibit small-world character, but scale-free degree
statistics can hardly be recognized. The results on the
probability degree statistics for the dual graphs of compact
urban patterns analyzed in accordance with the above described street identification
principle  are compatible
with that reported in \cite{Jiang2004}.
In general,
compact city patterns do not provide
us with sufficient data to conclude on the universality of
degree statistics.

\begin{figure}[ht]
 \noindent
\begin{center}
\epsfig{file=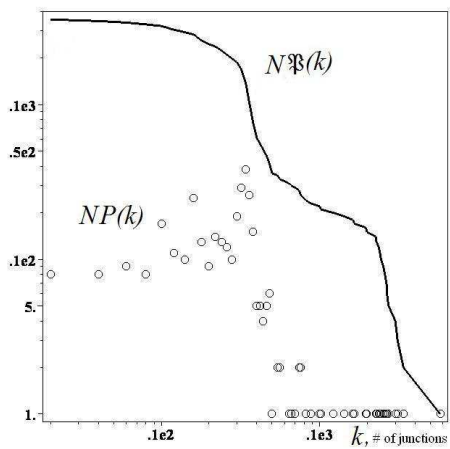, angle=0,width =6cm, height =6cm}
  \end{center}
\caption{\small The probability degree statistics for the dual graph of
 Manhattan (the log-log plot). The
number of
 streets in Manhattan, $NP(k)$,
 (displayed by circles) versus the number $k$ of
junctions they have. The solid line is for the cumulative
probability degree distribution (\ref{cumulativedistr}). }
\label{Fig1_09}
\end{figure}

The probability degree distributions for the
dual graph representations of the five compact urban patterns
mentioned in Tab.~1 has been studied by us in \cite{Volchenkov2007}.
To give an example,
we display in Fig.~\ref{Fig1_09}
the log-log plot is sketched
of the number of streets $NP(k)$ in Manhattan
versus the
number of their junctions $k$, where
\begin{equation}
\label{degdistr01}
P(k)=\Pr\left[i\in \mathfrak{G}|\deg(i)=k\right].
\end{equation}
 These numbers are
displayed by points.
The solid line is associated to cumulative
degree distribution, $N\mathfrak{P}(k),$
where
\begin{equation}
\label{cumulativedistr}
\mathfrak{P}(k)=\sum_{k'=k}^{N}P(k')
\end{equation}
 is the probability that the degree is {\it greater than
or equal} to $k$. The presentation of degree data by
 the cumulative degree distribution has an advantage over the degree
  distribution (\ref{degdistr01}), since it reduces the noise
  in the distribution tail, \cite{Newman2003}.
 It is remarkable that the
observed profiles are broad
indicating that a street in a compact
city can cross the different number of other streets,
 in contrast while
in a regular grid.
At the same time,
the distributions usually
 have a clearly noticeable maximum
corresponding to the most probable
number of junctions an average street has in the city.
The long right tail of the distribution
which could decay even faster then a power law
is due to just a few "broadways," embankments,
and belt roads crossing many more streets than
an average street in the city, \cite{Volchenkov2007}.
It has been suggested in
\cite{Figueiredo2007} that irregular shapes
 and faster decays in
the tails of degree statistics
indicate that
the connectivity distributions  is
{\it scale-dependent}.
As a possible reason for the
  non-universal behavior is that
 in
the mapping and descriptive procedures,
inadequate choices to determine the
 boundary of the maps or flaws
in the aggregation process can damage
the representation of very
long lines.
Being scale-sensitive in general,
the degree statistics of dual city graphs
 can nevertheless be approximately
universal within a particular range
of scales.

The scale-dependence of degree distributions
indicate that
the degree statistics alone
does not give us the enough information
 to reach a qualified conclusion on the structure of
  urban spatial
networks. Thus, the statistics of far-away neighbors
would be targeted to reduce the gap
providing us with a new
method for automatic structural
classification of cities.

\section{Acknowledgment}
\label{Acknowledgment}
\noindent

The work has been supported by the Volkswagen Foundation (Germany)
in the framework of the project: "Network formation rules, random
set graphs and generalized epidemic processes" (Contract no Az.:
I/82 418). The authors acknowledge the multiple fruitful
discussions with the participants of the workshop {\it Madeira
Math Encounters XXXIII}, August 2007, CCM - CENTRO DE CI\^{E}NCIAS
MATEM\'{A}TICAS, Funchal, Madeira (Portugal).


%
%

\end{document}